\pdfoutput=1
\documentclass[reprint,nofootinbib, amsmath,amssymb, aps, prl]{revtex4-1}

\usepackage{color}
\usepackage[pdftex]{graphicx}
\usepackage{dcolumn}
\usepackage{bm}
\usepackage{epstopdf}
\usepackage{natbib}

\begin{document}

\preprint{APS/123-QED}

\title{Embryonic lateral inhibition as optical modes:
 an analytical framework for mesoscopic pattern formation}

\author{Jose Negrete Jr$^{1}$}
 \email{jose.negretejr@epfl.ch}
\author{Andrew C. Oates$^{1}$}
\affiliation{$^{1}$\'{E}cole Polytechnique F\'{e}d\'{e}rale de Lausanne (EPFL),  CH-1015 Lausanne, Switzerland\\}




\date{\today}

\begin{abstract}
Cellular checkerboard patterns are observed at many developmental stages of embryos.
We study an analytically tractable model for lateral inhibition and show that a coupling coefficient
with a negative value is sufficient to obtain noisy or periodic checkerboard patterns.
We solve the case of a linear chain of cells explicitly and show that noisy anti-correlated 
patterns are available in a post-critical regime $(\epsilon_c < \epsilon < 0)$. In the sub-critical regime 
$(-\infty < \epsilon \leq \epsilon_c)$ a periodic and alternating steady state is available, 
where pattern selection is determined by making an analogy with the optical modes of phonons. For cells 
arranged in a hexagonal lattice, the sub-critical pattern can be driven into three different states: two of those states are periodic checkerboards and a
third in which both periodic states coexist.
\end{abstract}

\pacs{Valid PACS appear here}
\maketitle


\textit{Introduction.-} Pattern formation in living tissue is an emergent property
that arises from cell signaling. In his seminal work Alan Turing proposed
that the anatomical structure of an embryo is determined
by self-organised chemical patterns~\cite{Turing_1952}. In his theory the interplay between reactions
and diffusion of two chemical species give rise to self-organized periodic structures. 
Although this model is difficult to implement in practice with the restrictions
proposed by Turing~\cite{Castets_1990, Ouyang_1991}, variants of this model system have been extensively studied 
theoretically~\cite{Cross_2009} and have been successfully used to describe the patterns of living systems at 
the tissue level~\cite{Kondo_1995, Yamaguchi_2007, Raspopovic_2014}.

There is a subclass of patterns in living systems where the characteristic 
length scale $(\lambda)$ is the size of a single cell (figure 1 a). These are fine-grained patterns
where the protein expression levels vary abruptly and regularly between 
cells, reminiscent of checkerboards (figure 1 a). In this case the reaction-diffusion
mechanism from Turing cannot describe their formation,
as its characteristic length scale corresponds to a wavenumber equal to infinity
in a continuum description. These patterns appear in several examples of cellular 
tissues such as in the arrangement of photo receptor cells in the eye~\cite{Gavish_2015},
sensory hair cells in the auditory epithelium~\cite{Togashi_2011, Adam_1998}, 
and sensory bristles in the fly thorax~\cite{Corson_2017} among others.
Lateral inhibition has been proposed as a basic mechanism whereby a cell inhibits the expression levels of 
a protein in its neighbouring cells. The Delta-Notch signaling system is a pathway 
frequently used by different species to create these fine-grained tissue patterns~\cite{Artavanis_1999}.

Collier et al.~\cite{Collier_1996} pioneered theoretical work in lateral inhibition, 
analysing a spatially discrete model for Delta-Notch signaling that reproduced
periodic and fine-grained patterns. Since then, extensions of this model (or similar ones) have incorporated
different effects observed in experiment, such as self regulation (cis-interactions)~\cite{Sprinzak_2011}, 
state dependent coupling strength~\cite{Formosa_2014}, time delay in signaling~\cite{Veflingstad_2005, Momiji_2009} 
and cis-interactions with time delay~\cite{Glass_2016}. These models
are difficult to analyse analytically and they rely on numerical simulations
to determine how patterns emerge. In this work we propose a generic 
model for lateral inhibition that is analytically tractable, allowing us to understand
pattern selection for this class of systems. The model consists of 
the following equation

\begin{figure}
\includegraphics[width=0.5\textwidth]{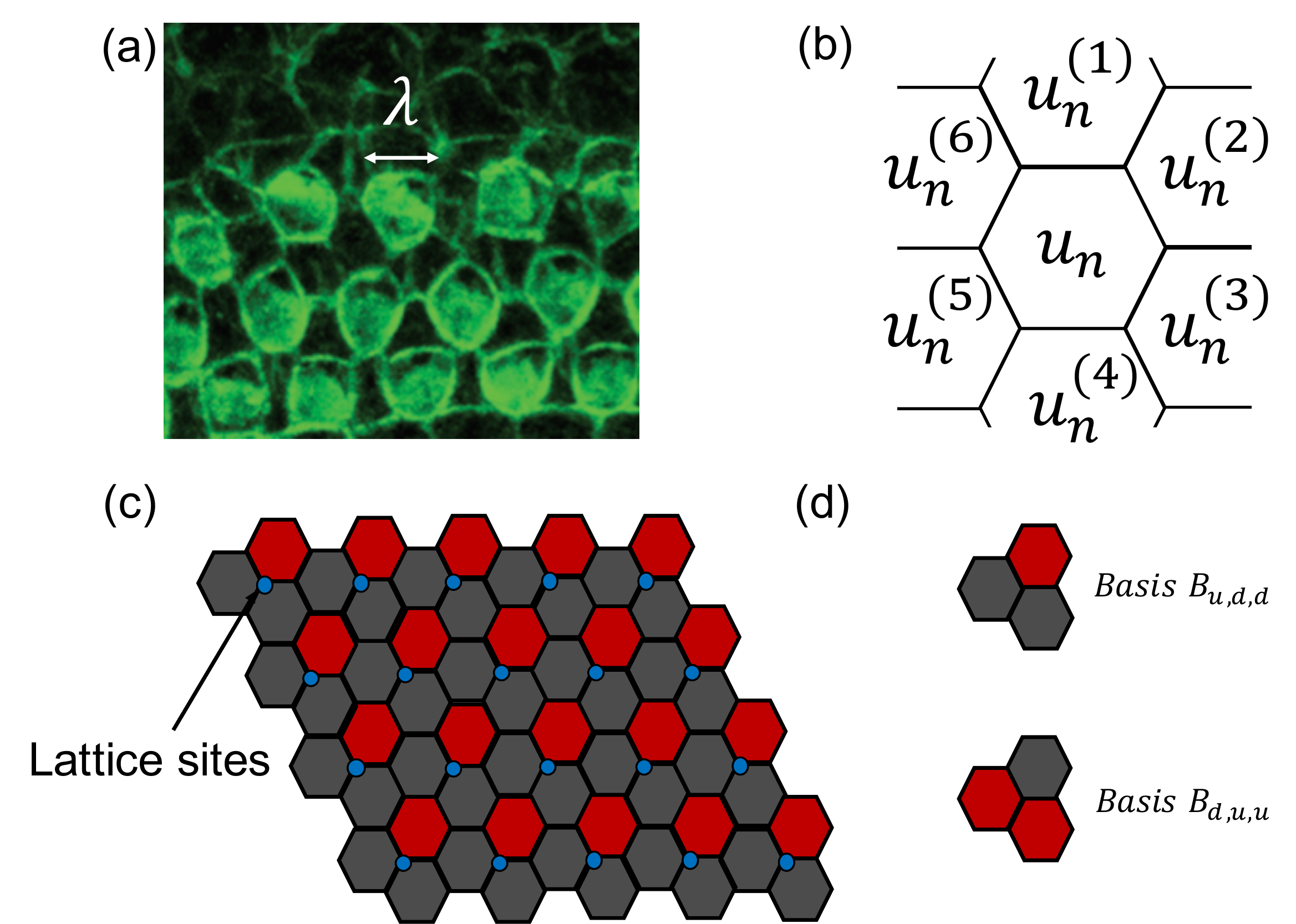}
\caption{Periodic fine grained patterns, (a) example from the auditory epithelium with a green fluorescent marker for
actin modified from ~\cite{Togashi_2011},  $\lambda$ depicts the characteristic wavelength of the pattern,
(b) depiction of the notation used for Equation (1), (c) pattern conformed by an hexagonal
lattice, where at each lattice point there is a basis structure with three cells and
(d) the possible basis structures for a periodic pattern.}
\label{Lattice}
\end{figure}

\begin{equation}
\frac{d u_{n}}{dt} = \underbrace{-u_{n} - \gamma u_{n}^3+\Omega_{n}}_{cis-regulation} + \underbrace{\epsilon \sum\nolimits_{i = 1}^N u_{n}^{(i)}}_{trans-regulation}
\label{model_2D}
\end{equation}

\noindent where the state of each cell placed in a lattice is represented by the variable $u_{n}$.
The model describes the effects of \textit{cis-interactions} where $\Omega_{n}$ is an 
internal component that influences the production of $u_{n}$ and $\gamma > 0$ is the 
strength of nonlinear degradation. The cells are influenced by the state of their $N$
nearest neighbours $u_{n}^{(i)}$ (\textit{trans-regulation}) with coupling strength 
$\epsilon$ (figure 1 b).

In this work we determine the conditions necessary to
obtain noisy anticorrelated patterns and periodic alternating
patterns for $u_n$ given the signaling strength $\epsilon$ and the profile
of the internal production rate $\Omega_{n}$. These lateral inhibition patterns
appear only when the value of $\epsilon$ is negative. If we set $\Omega_n = 0$ 
in equation (1), there is a sub-critical regime $(\epsilon\leq\epsilon_c)$
where the uniform state $u_n = 0$ is unstable and a post-critical 
regime $(\epsilon>\epsilon_c)$ where $u_n = 0$ is stable.
In the sub-critical regime we consider a spatially uniform profile
for $\Omega_n$ while in post-critical regime we consider spatially 
uncorrelated and time-constant values of $\Omega_n$.

The case of a one dimensional chain of cells is explicitly solved, and these 
results helps us to analyse the case of cells placed in an hexagonal lattice (figure 1 c). 
By making an analogy to lattice (phonon) vibrations from solid state physics~\cite{Kittel}, 
we determine a potential function that determines the possible sub-critical states 
of the system. In cells arranged in an hexagonal lattice, these states 
lie on a periodic lattice (blue points in figure 1 c), where given $\Omega_n$
we obtain: 1) a periodic pattern filled with the basis structure denoted $B_{u,d,d}$ (figure 1 d),
2) a pattern filled with the basis $B_{d,u,u}$ (figure 1 d), or 3) a state where 
both basis $B_{u,d,d}$ and $B_{d,u,u}$ coexist.

\textit{Linear reponse of a chain of cells.-} We consider a one dimensional
chain of cells signaling each other following Equation (1).
For this example, we define a cell in the middle of the chain
with index $n = 0$. First consider the linearized version
of Equation (1) around $u_n = 0$, which is given by

\begin{equation}
\frac{d u_n}{dt} \approx -u_n + \epsilon [ u_{n-1} + u_{n+1}] + \Omega_n
\label{model_linearized}
\end{equation}

\noindent As a first step to charaterize its dynamics we study the 
linear response function of the chain to the static input $\Omega_{n} = \Omega_o \delta_{n0}$,
where $\delta_{nn'}$ corresponds to the Kronecker delta.
The linear response of an infinite chain is given by the following expression 

\begin{equation}
u_{n} = \frac{\Omega_o}{\sqrt{1-4 \epsilon^2}}\Big[ \frac{1 - \sqrt{1 - 4 \epsilon^2}}{2 \epsilon} \Big]^{|n|}
\label{analytical}
\end{equation}

\begin{figure}
\includegraphics[width=0.525\textwidth]{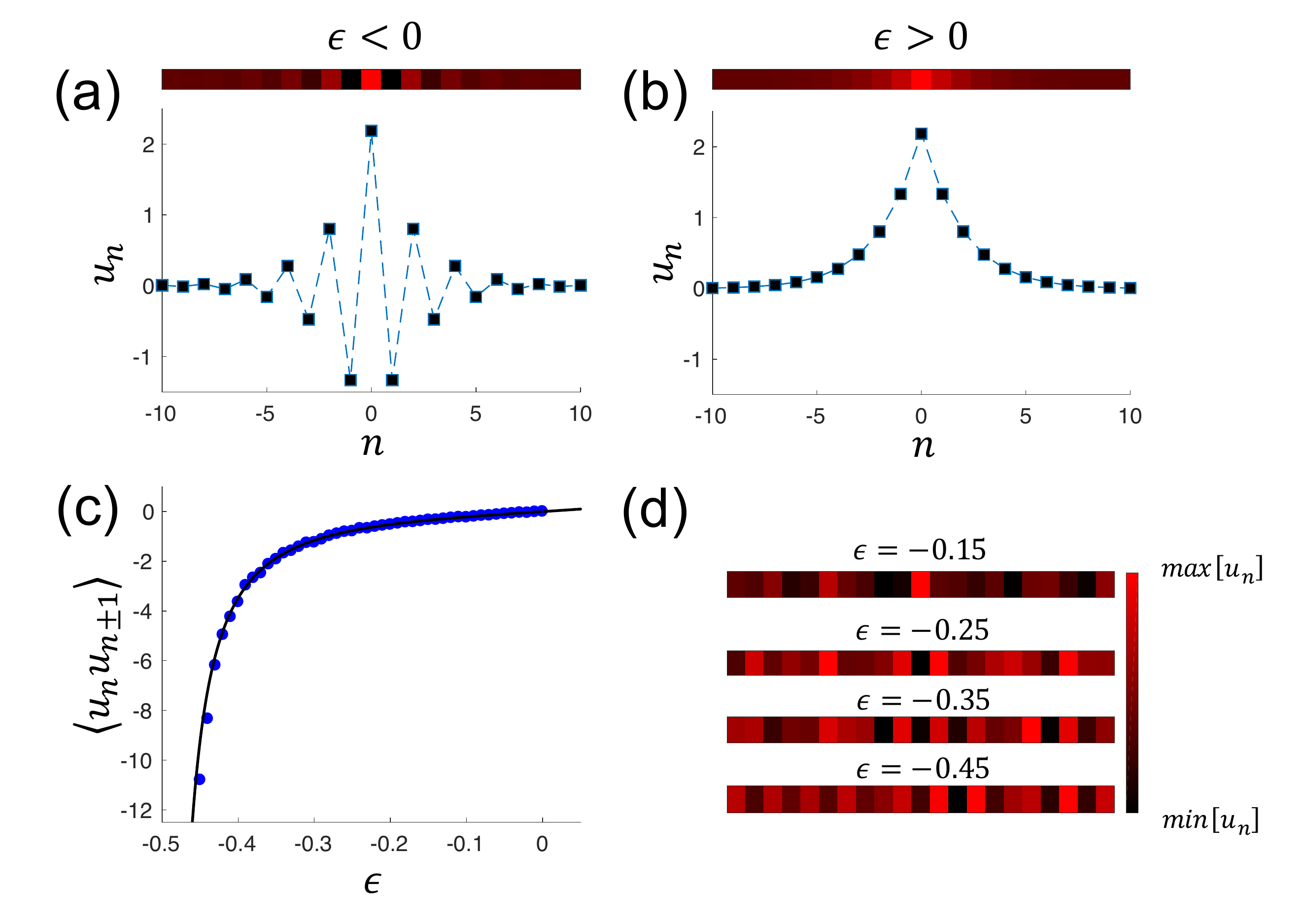}
\caption{Impulse response for a linear chain of cells. We set
$\Omega_n = \Omega_o \delta_{00}$ where (a) corresponds to $\epsilon<0$
and b) to $\epsilon>0$ for a finite chain of 20 cells. Response to random input $\Omega_n$ where 
(c) shows the next neighbour correlations $<u_n u_{n\pm1}>$ as a function of $\epsilon$, symbols correspond
to numerical simulations and solid curve was obtained analytically (Equation~(\ref{correlations})) and
(d) are examples of noisy anti-correlated patterns as a function of $\epsilon$.}
\label{Response}
\end{figure}

\noindent (see supplement for details~\cite{Supplement}). 
Note in equation (3) there is a term proportional to $\epsilon^{-|n|}$,
therefore if $\epsilon<0$ the values of $u_n$ alternate between positive 
and negative values (figure~\ref{Response} a, lateral inhibition), 
while for $\epsilon>0$ the values of $u_n$ decay smoothly 
(figure~\ref{Response} b, lateral induction).

We next calculate the correlations in a chain
of $u_n$ now considering an stochastic input $\Omega_n$ characterized by

\begin{equation}
\Big< \Omega_n  \Omega_{n'} \Big> = \sigma^2_{\Omega} \delta_{nn'}
\end{equation}

\noindent The linear response function is used to calculate the 
correlation between nearest neighbors~\cite{Supplement},
which is given by

\begin{equation}
\Big< u_n u_{n \pm 1} \Big> \approx  \frac{2 \sigma^2_{\Omega}}{\epsilon} \Big( 1 - \frac{1}{\sqrt{1 - 4 \epsilon^2}} \Big)^2
\label{correlations}
\end{equation}

\noindent The correlations are zero for $\epsilon = 0$ and they
diverge to $-\infty$ for $\epsilon = -1/2$, this is the same value found in Collier et al.~\cite{Collier_1996} 
by linear stability analysis. The divergence in the correlation coefficient
is characteristic of a critical point~\cite{Scheffer_2009}, while its 
negative sign denotes that the neighbors are anti-correlated. 
At this critical point $(\epsilon = \epsilon_c)$ the uniform state 
$u_n = 0$ becomes unstable. There is good agreement between numerical and 
the analytical approximation given by Equation~(\ref{correlations}) (see 
Figure~\ref{Response} c). Examples of anti-correlated noisy patterns 
in Figure~\ref{Response} (d) show that as $\epsilon$ decreases, the probability 
of having alternating neighbours increases.

\textit{Analogy to optical modes.-} The results shown in Figure 2 (a)
and (b) are reminiscent of the optical and acoustic modes of phonons from solid 
state physics~\cite{Kittel}. Phonons are vibrations in a periodic crystal. The crystal is composed
by a set of lattice sites, where at each site there is a molecule with several atoms
called the basis (also termed as the crystal unit cell). If the basis is diatomic, two vibration modes are allowed.
In the optical mode the position of atoms alternate between positive and 
negative values between neighbors, and in the acoustic mode this alternation
does not happen~\cite{Kittel}. The linear response of $u_n$ for $\epsilon<0$ (figure~\ref{Response} a) 
is analogous to the optical mode and for $\epsilon>0$ to the acoustic mode (figure~\ref{Response} b).

\begin{figure}
\hspace*{-0.30cm}
\includegraphics[width=0.45\textwidth]{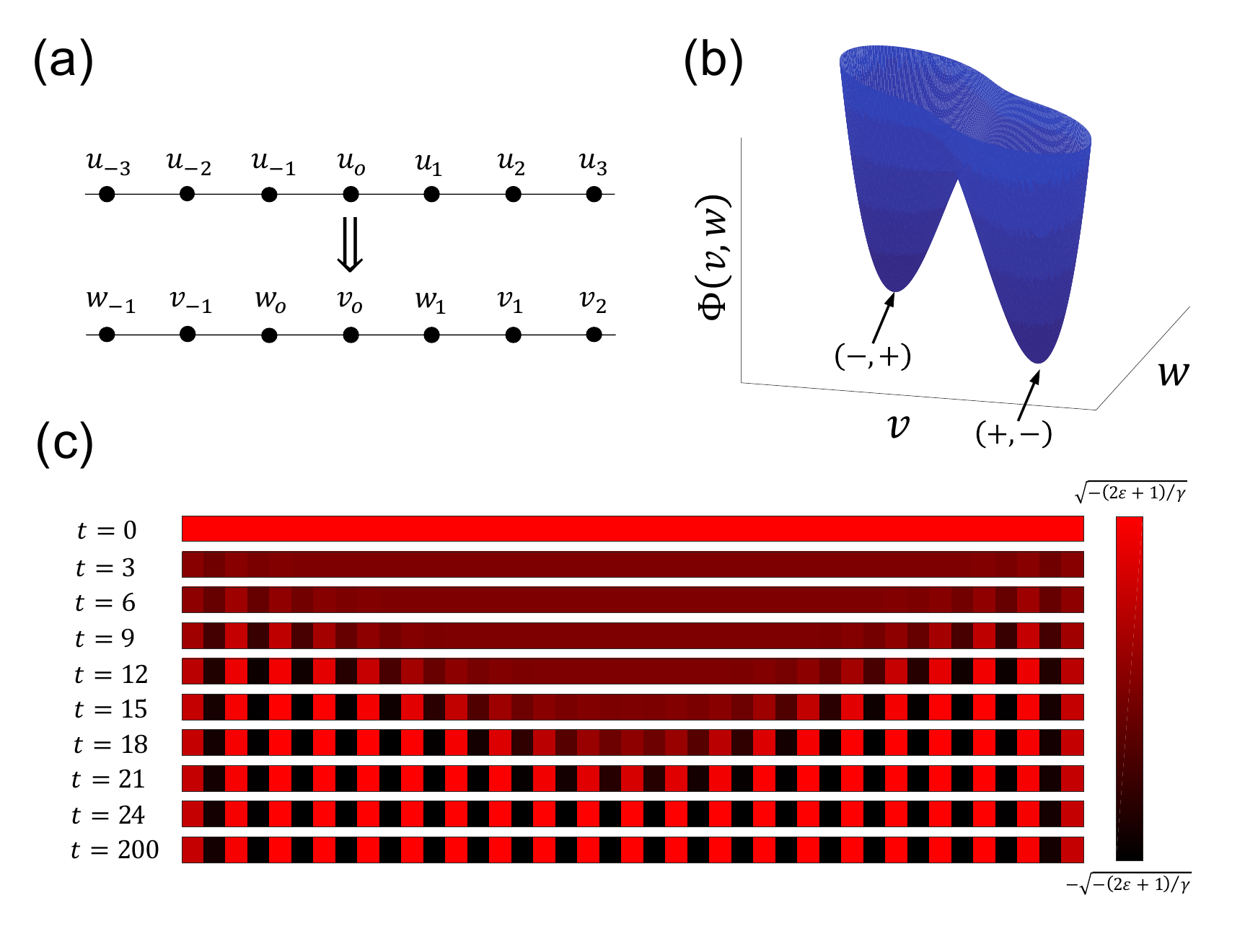}
\caption{Potential well for a linear chain of cells. (a) Variable transformation underlying
calculation of potential well. (b) An example of a potential 
$\Phi(v,w)$ with $\epsilon<1/2$, the potential has minima at $(+,-)$ 
(positive $v_s$ and negative $w_s$) and $(-,+)$ (negative $v_s$ 
and positive $w_s$). (c) Time evolution of a numerical simulation
of a linear chain using Equation (1).}
\label{Nonlinear_chain}
\end{figure}

This analogy to the optical modes
is useful to analyze the sub-critical state $(\epsilon<\epsilon_c)$.
We start by setting a uniform $\Omega_n = \Omega_o$
 in equation (1). We divide the chain of $u_n$ into $s$ lattice sites and at each site
there is a basis with two components $v_s$ and $w_s$. The components $v_s$
corresponds to $u_n$ with $n = 0$ and even values of $n$, and the remainder
corresponds to $w_s$ (figure~\ref{Nonlinear_chain} a). In this notation equation~(1) for the one dimensional chain becomes

\begin{equation}
\frac{d v_s}{dt} = -v_s - \gamma v_s^3 + \epsilon [ w_{s-1} + w_{s}] + \Omega_o
\label{v_sites}
\end{equation}

\begin{equation}
\frac{d w_s}{dt} = -w_s - \gamma w_s^3 + \epsilon [ v_{s} + v_{s+1}] + \Omega_o
\label{w_sites}
\end{equation}

We rewrite equations~(\ref{v_sites}) and (\ref{w_sites}) in the 
continuum approximation as 
 
 \begin{equation}
\frac{\partial v}{\partial t} = 2\epsilon w - v - \gamma v^3 + \epsilon \Delta x^2 \frac{\partial^2 w}{\partial x^2} + \Omega_o
\label{v_sites_cont}
\end{equation}

\begin{equation}
\frac{\partial w}{\partial t} = 2\epsilon v -w - \gamma w^3 + \epsilon \Delta x^2 \frac{\partial^2 v}{\partial x^2} + \Omega_o
\label{w_sites_cont}
\end{equation}

\noindent where $x$ corresponds to the position in the line tissue and $\Delta x$
is the distance between the lattice points. Assuming
a uniform stationary state for both $v_s$ and $w_s$
then we define a potential

\begin{equation}
\Phi(v,w) =  - \Omega_o(v+w) +\frac{v^2+w^2}{2} -2 \epsilon v w + \gamma \frac{v^4+w^4}{4} 
\label{Potential_line}
\end{equation}

\begin{figure}
\hspace*{-0.30cm}
\includegraphics[width=0.5\textwidth]{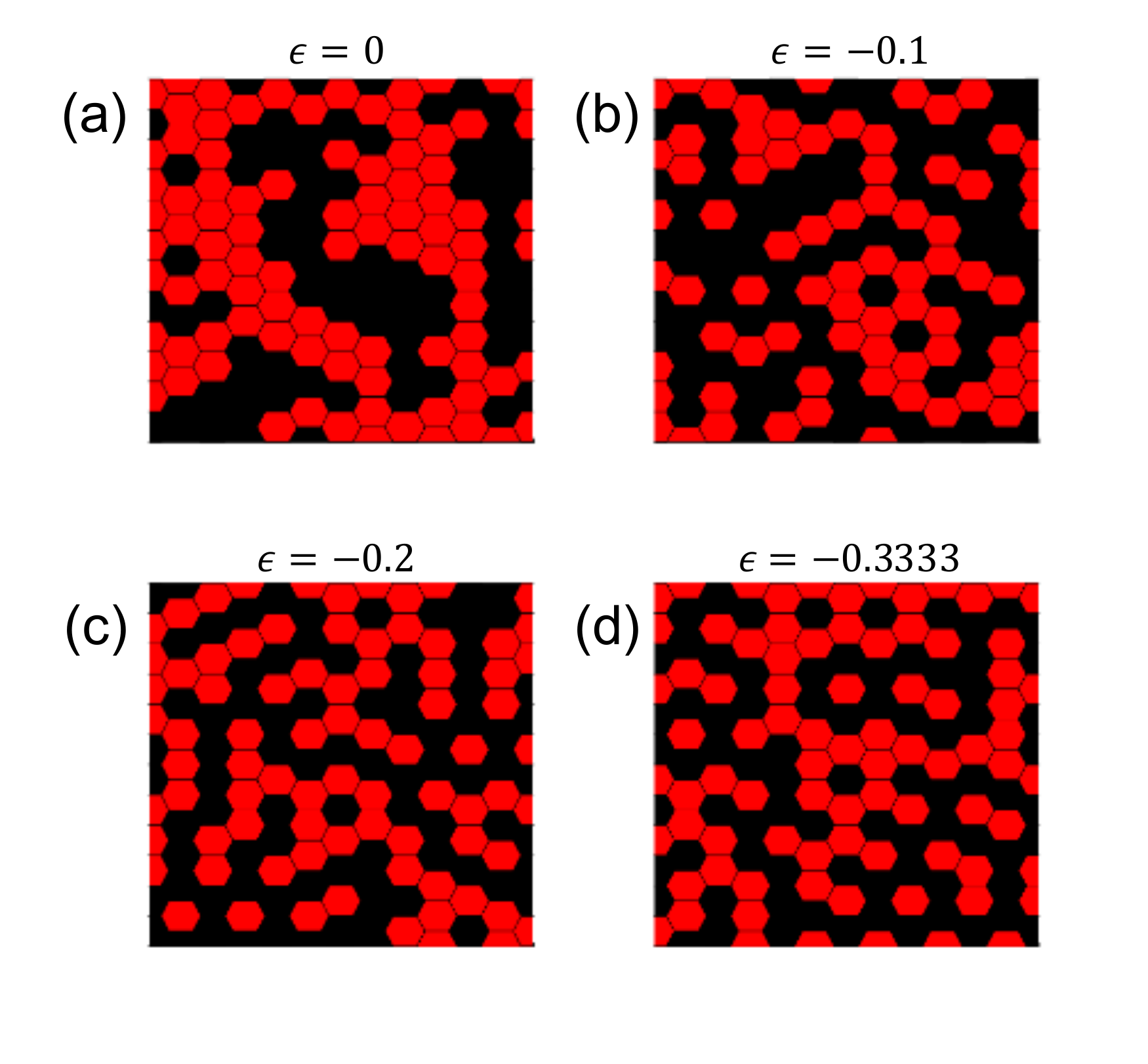}
\caption{Increase in cell anti-corerelation with increasing coupling strength.
Stochastic patterns given by $f_n = 2 \theta(u_n)-1$ generated with noisy input 
$\Omega_n$, black corresponds to $f_n = -1$ and red to $f_n = 1$.}
\label{Stochastic}
\end{figure}

\noindent for $\epsilon <\epsilon_c$ this is a double well potential (figure~\ref{Nonlinear_chain} b). 
One minimum of this potential we denote as the state $(+,-)$ where 
$v_s = c_o$ and $w_s = -c_o$. The other minimum we denote as $(-,+)$ 
where $v_s = -c_o$ and $w_s = c_o$. From this potential we expect that the steady 
states of Equation (1) consists of anti-correlated neighbors. We test this prediction 
with a numerical simulation of equation (1). We set $\Omega_o =0$ and the uniform initial 
condition to $u_n = \sqrt{-(2\epsilon+1)/\gamma}$ (figure~\ref{Nonlinear_chain} (c) at $t=0$). 
This initial state is unstable and the values decrease towards $u_n = 0$ ($t = 3$).  A wave 
emerges from the edges of the chain and travels towards the center, reminiscent of 
waves observed in the Collier model~\cite{Plahte_2006}. The values of $u_n$ intercalate between values close 
to $\pm c_o$, in this regime $c_o = \sqrt{-(2\epsilon+1)/\gamma}$. At long times the pattern remains in this alternating state 
{(figure~\ref{Nonlinear_chain} (c) at $t=200$). Thus there is agreement between the predictions of the potential and the steady state
of the system.

\begin{figure*}
\includegraphics[width=1.0\textwidth]{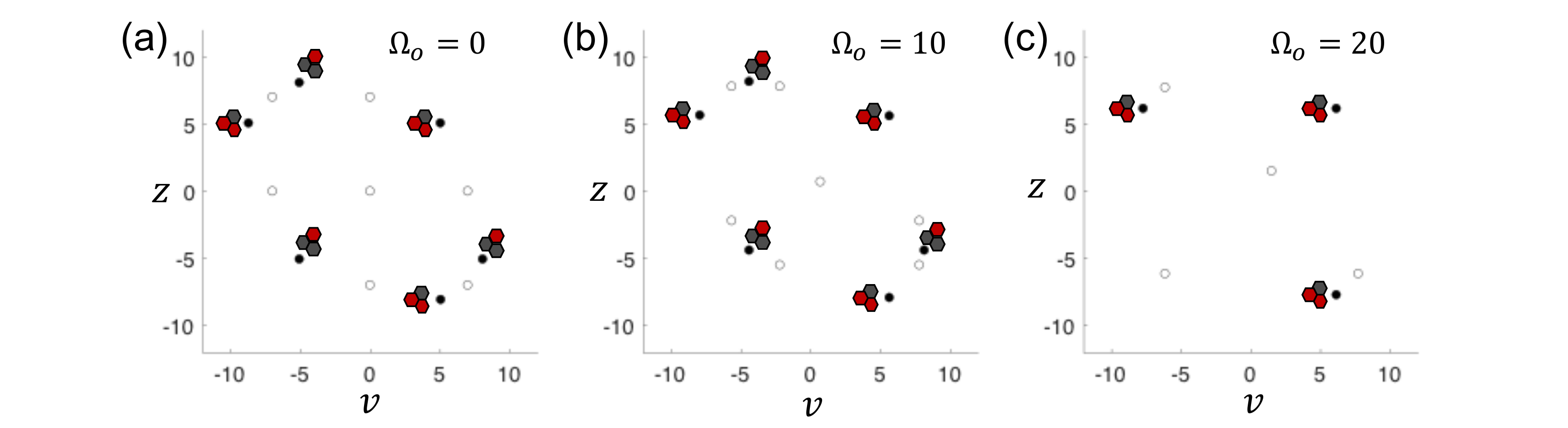}
\caption{Stable states (black symbols) and metastable states (white symbols)
of the potential $\Phi(v,w,z)$ given by Equation~(\ref{Potential_tissue}) projected onto the $v-z$ axis for the
parameters $\gamma = 0.1$ and $\epsilon = -2$.}
\label{Points}
\end{figure*}

\begin{figure}
\includegraphics[width=0.4\textwidth]{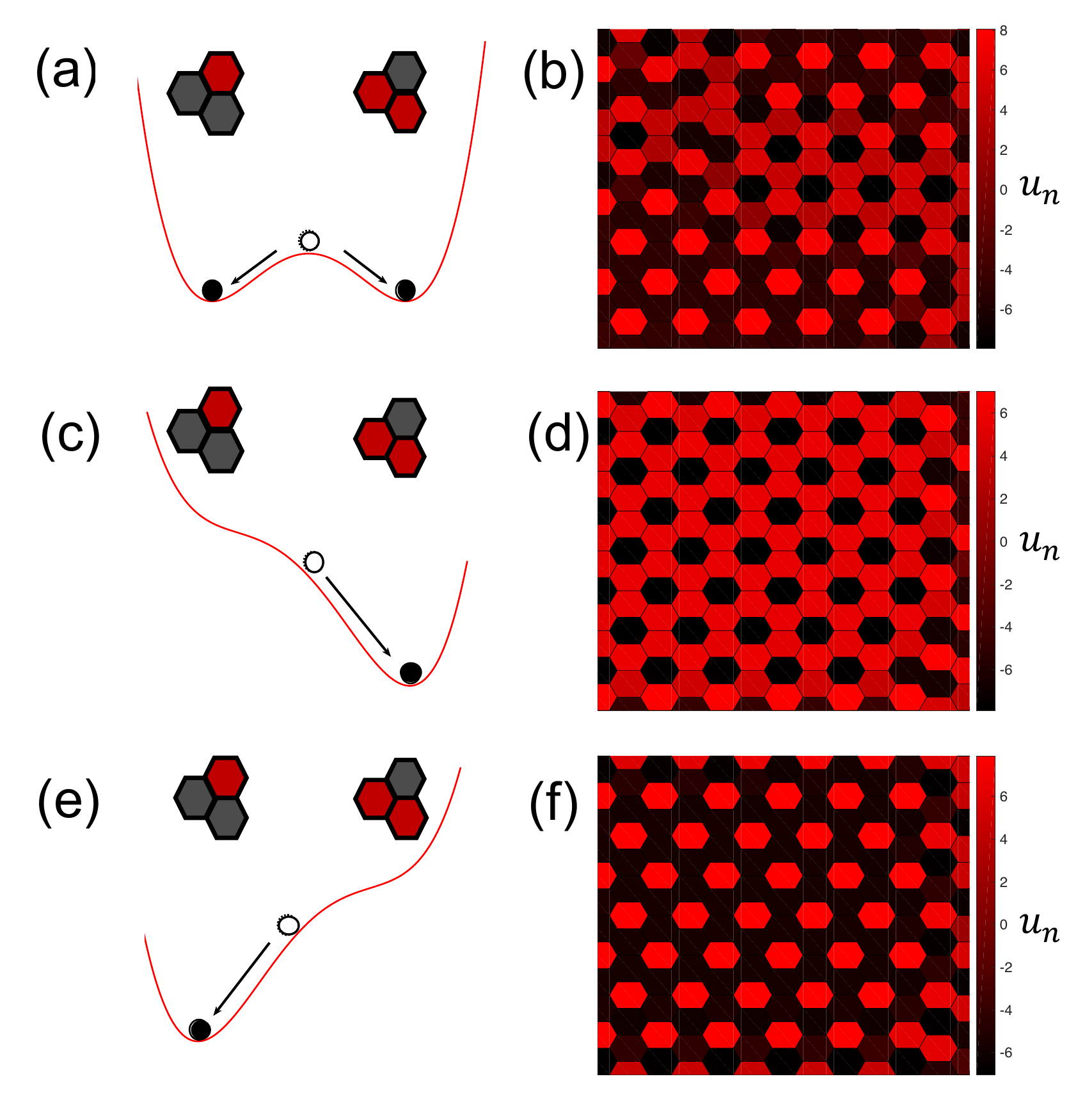}
\caption{Pattern selection and conceptual double well potential, (a) double well potential where the minima
represents the basis $B_{u,d,d}$ and $B_{d,u,u}$ and (b) its corresponding pattern where both states coexist.
For $\Omega_o> \Omega_c$ the potential (c) has a single minima at $B_{d,u,u}$ and its corresponding pattern
is shown in (d), while for $\Omega_o < - \Omega_c$ the potential (e) has a single minima at $B_{u,d,d}$
and its corresponding pattern is shown in (f).}
\label{Patterns}
\end{figure}

\textit{Cells in an hexagonal lattice.-} Typical fine-grained pattern formation
occurs in epithelial tissues, single layers of cells. As a good approximation of
these tissues, we analyze the case of cells placed in an hexagonal lattice.
Using the observations of a linear chain of cells we analyze
the response $u_n$ to a random input $\Omega_n$ in epithelial tissues. 
Figure~\ref{Stochastic} shows the projection $f_n = 2\theta(u_n)-1$, where $\theta$
corresponds to the Heaviside theta, for} different steady states as a function $\epsilon$.
As in the example of the previous section, we see that the likelihood of being
surrounded by anti-correlated neighbors increases as the value of $\epsilon$ decreases.

We proceed by analyzing the sub-critical regime of equation (1) by 
dividing $u_n$ into different components as in the previous example. Again we consider a 
spatially uniform $\Omega_n = \Omega_o$. Defining a basis
with two variables as in the linear chain, and rewriting 
Equation (1) we notice that the interactions are asymmetric~\cite{Supplement}.
Therefore we rewrite Equation (1) in terms of an hexagonal lattice with lattice sites $s$
(see figure 1 c) and place at each lattice site a basis with three components 
$v$, $w$ and $z$. We derive a potential well for this case ~\cite{Supplement}.

\begin{equation}
\begin{split}
\Phi(v,w,z) =  - \Omega_o(v+w+z) +\frac{v^2+w^2+z^2}{2}\\
 -3 \epsilon [v w + v z +w z] + \gamma \frac{v^4+w^4+z^4}{4} 
 \end{split}
\label{Potential_tissue}
\end{equation}

The state of the tissue becomes unstable at $\epsilon_c = -1/3$~\cite{Collier_1996}, in this case we find 
the minima of this potential and their stability numerically. 
Figure~\ref{Points} shows the position of the stable (black points) and the
metastable states (white points) projected onto the $v-z$ axis from the potential given by equation~(\ref{Potential_tissue}), as a
function of $\Omega_o$ for $\epsilon < \epsilon_c$~\cite{Supplement}. When $\Omega_o = 0$ there are six stable states,
three correspond to the basis $B_{u,d,d}$ (figure~\ref{Points} a) mentioned in the introduction 
(see figure~\ref{Lattice} d). The other stable states corresponds to the basis $B_{d,u,u}$. 
A simplified picture of this potential is shown in Figure~\ref{Patterns} (a) as a double well potential where 
each minima corresponds to a base. In figure~\ref{Patterns} (b) 
we show a numerical simulation of Equation (1) in this regime initiated 
from uniform initial conditions. The final pattern has two coexisting regions, 
one filled with the basis  $B_{u,d,d}$ and the other filled with $B_{d,u,u}$.

If we increase further the value of $\Omega_o > 0$ (figure~\ref{Points} b), 
the points indicating the metastable states move closer to 
the points indicating the basis $B_{u,d,d}$. After crossing a critical value 
for $\Omega_c$, the stable points corresponding to $B_{u,d,d}$ disappear 
by colliding with a metastable point. In the simplified picture shown
in Figure~\ref{Patterns} by increasing the parameter $\Omega_o$ the 
double well potential (Figure~\ref{Patterns} a) becomes a single well
potential (Figure~\ref{Patterns} c). The single minima corresponds to
the basis $B_{d,u,u}$. Numerical simulations in this regime results 
in a pattern filled with the basis $B_{d,u,u}$ (figure~\ref{Patterns} d). 
If we repeat the same procedure as before but decreasing the control
parameter $\Omega_0 < - \Omega_c$, then the potential has a single
well with the minima corresponding to $B_{u,d,d}$ (figure~\ref{Patterns} e). Simulations in this 
regime shows a pattern filled with $B_{u,d,d}$ (figure~\ref{Patterns} f). 
We note that some final patterns can contain defects depending on initial 
conditions~\cite{Supplement}.

\textit{Conclusions.-}  In this work we have proposed an analytically
tractable model for lateral inhibition. We have shown how fine-grained checkerboard
patterns depend on the model parameters. By making an analogy with 
optical modes from lattice phonons we are able to determine
how fine-grained patterns are selected. Thus our
model is a generic description of lateral inhibition that explains the
transitions observed in more detailed models as the 
ones in~\cite{Collier_1996, Sprinzak_2011, Formosa_2014, Veflingstad_2005, Momiji_2009,Glass_2016}.

Our model brings a new framework to studying and understanding pattern selection in 
both natural and synthetic biological fine-grained systems. Particularly attractive experimental 
scenarios for testing these ideas are in vitro cellular systems with synthetic intercellular 
signaling pathways such as recently reported by Matsuda et al~\cite{Matsuda_2014}. Disrupting Delta-Notch 
communication between cells in this system using a chemical inhibitor yielded a uniform 
expression of Delta across the tissue. Removal of the inhibitor allowed Delta-Notch signaling 
to initiate and caused the evolution of a fine-grained pattern of Delta expression. This 
mimics our model's behaviour when the coupling strength is modulated from $\epsilon = 0$ 
to $\epsilon < \epsilon_c$. This and similar experimental systems~\cite{Toda_2018, Glass_2018} 
also raise the possibility of controlling the initial expression levels of Delta, or an equivalent signal, 
which would be analogous to modulating $\Omega_o$ in the model.

Finally, our analysis shows that pattern formation by cell signalling
has two regimes. A regime where the steady state is composed by 
acoustic modes, this is the reaction-diffusion regime scenario
proposed by Turing, and a regime composed by optical modes 
that corresponds to the fine-grained patterns of lateral inhibition. 

\textit{Acknowledgements.- We would like to thank
Marta Iba\~{n}es and Jos\'{e} M. Sancho for sharing their 
unpublished manuscript with a similar model. Their analysis and
observations are complementary to our work.}

\nocite{*}

\bibliography{Bibliography_Lat}
   
\end{document}